\documentclass[runningheads]{llncs}
\usepackage{hyperref}
\usepackage[T1]{fontenc}
\usepackage{graphicx}
\usepackage{color}

\usepackage{xcolor}
\usepackage{soul}
\usepackage{tabularx,booktabs} 
\usepackage{algorithm}
\usepackage{algpseudocode}
\newcolumntype{P}[1]{>{\centering\arraybackslash}p{#1}}
\newcolumntype{M}[1]{>{\centering\arraybackslash}m{#1}}
\newcommand{\repeatthanks}{\textsuperscript{\thefootnote}}

\begin{document}
\title{Seamless Iterative Semi-Supervised Correction of Imperfect Labels in Microscopy Images}
%
\titlerunning{SISSI: Seamless Iterative Semi-Supervised Correction of Imperfect Labels}
%
\author{Marawan Elbatel\thanks{Co-first authors} \and
Christina Bornberg\repeatthanks \and Manasi Kattel \and Enrique Almar \and Claudio Marrocco \and Alessandro Bria}
\authorrunning{Elbatel  et al.}
%
\institute{University of Cassino and Southern Lazio, Cassino, Italy \email{marawan.elbatel@studentmail.unicas.it
}
}


%
\maketitle              
\begin{abstract}


In-vitro tests are an alternative to animal testing for the toxicity of medical devices. Detecting cells as a first step, a cell expert evaluates the growth of cells according to cytotoxicity grade under the microscope. Thus, human fatigue plays a role in error making, making the use of deep learning appealing.
Due to the high cost of training data annotation, an approach without manual annotation is needed.
We propose \textit{Seamless Iterative Semi-Supervised correction of Imperfect labels~(SISSI)}, a new method for training object detection models with noisy and missing annotations in a semi-supervised fashion. Our network learns from noisy labels generated with simple image processing algorithms, which are iteratively corrected during self-training.
Due to the nature of missing bounding boxes in the pseudo labels, which would negatively affect the training, we propose to train on dynamically generated synthetic-like images using seamless cloning. 
Our method successfully provides an adaptive early learning correction technique for object detection. The combination of early learning correction that has been applied in classification and semantic segmentation before and synthetic-like image generation proves to be more effective than the usual semi-supervised approach by $>15\%$ AP and $>20\%$ AR across three different readers. Our
code is available at \url{https://github.com/marwankefah/SISSI}.

\end{abstract}

\section{Introduction}

Testing medical devices with animals have a long tradition according to ISO 10993 \cite{Anderson2016iso}. Since 2017 the ISO 10993 has gradually evolved towards implementing alternative test methods. One of the in-vitro methods is the testing of cytotoxicity, described in the ISO 10993-5 \cite{International2009iso}. Cell experts analyze cell growth of a fibroblast cell line such as L929 with the help of a microscope. The acceptance criteria for medical devices is 50\% of dead cells (grade 2 criteria). If there are more than 50\% dead cells, the medical device is not allowed to enter the market.
\par
In this context, deep learning can serve as a second opinion since human error in the workplace is costly and dependent on the level of fatigue; the greater the level of fatigue, the higher the risk of errors occurring.
Especially in the borderline cases of grade 2, the cell expert needs to be able to obtain a second opinion that is independent of human fatigue. Deep learning has shown substantial benefits in different life science and pharma applications such as chemo-informatics, computational genomics, and biomedical imaging such as cell segmentation \cite{Siegismund2018developing} and seems to be a promising supplement to cytotoxicity grading. In the first instance, cells need to be detected, and in future work, an intuitive way of classifying cells into dead or alive needs to be found.

\begin{figure}[t]
    \centering
    \includegraphics[width=\textwidth]{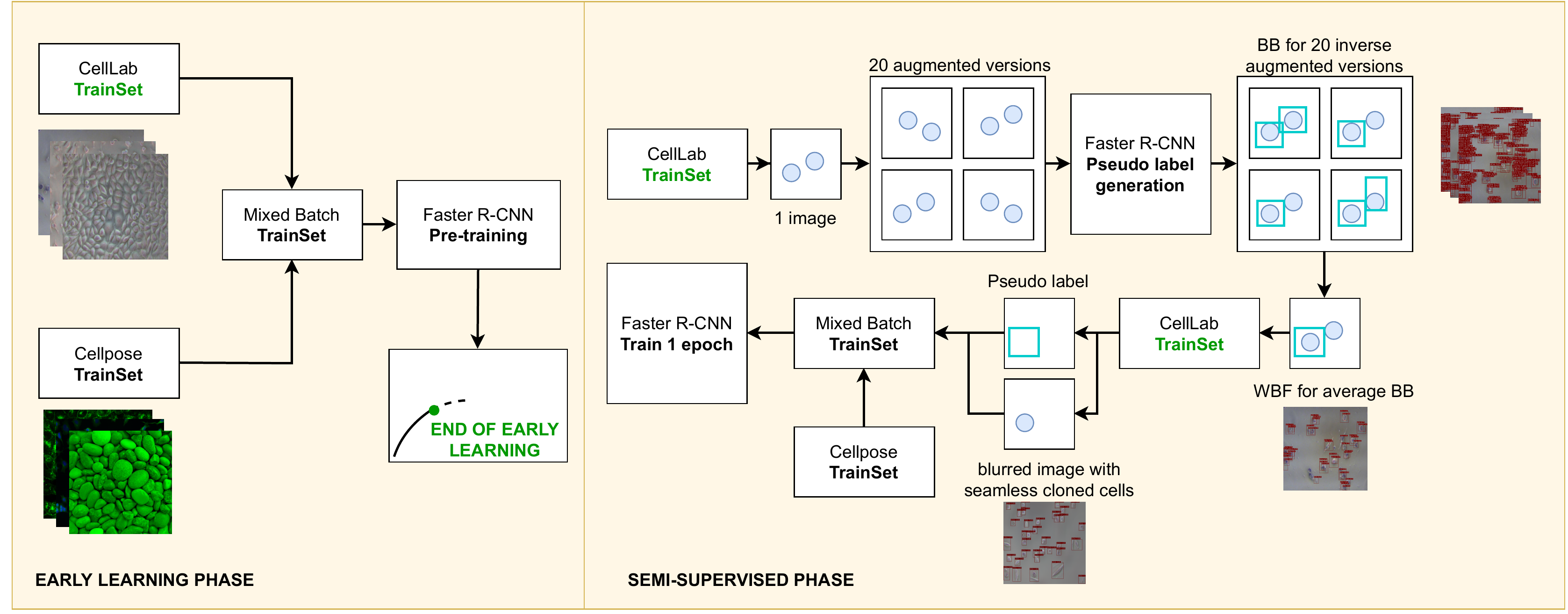}
    \caption{Overall scheme of SISSI framework.}
    \label{fig:sissi}
\end{figure}

\par
When dealing with imperfect datasets, problems including (partly) missing, inaccurate, or wrong labels arise. To handle imperfect datasets in object detection/segmentation tasks, one can leverage unlabelled (self/semi-supervised) or external labelled (transfer learning) data, regularise training, learn with class labels, and revisit loss functions (sparse/noisy labels) \cite{Tajbakhsh2020}.
\section{Related Works}
\par Previous work has studied imperfect datasets, including semantic segmentation, instance segmentation, and object detection \cite{Wang2022unreliable,Xiong2022UnetCHT,Li2022}.
\cite{Wang2022unreliable}~propose a pipeline for semantic semi-supervised segmentation that separates pixels of a pseudo labelled image into reliable and unreliable.
\cite{Liu2022adaptive}~propose Adaptive Early Learning Correction (ADELE) for semantic segmentation, with a supervised early-learning phase and subsequently a label correction phase. \cite{Lyu2021}~propose a label mining pipeline for missing annotations using co-teaching for instance segmentation.
\cite{Xiong2022UnetCHT}~propose to generate masks with the Circle Hough Transform (CHT) and iteratively create pseudo labels with self-training for images where CHT failed. 
\cite{Zhang2020recalibrationloss}~propose to use a background calibration loss inspired by focal loss for object detection with missing annotations.
\cite{Li2022}~propose only annotating one instance per category in an image and iteratively generating pseudo-labels.
\cite{Gao_2019_ICCV}~propose an object detector to handle noisy labels, masking the negative sample loss in the box predictor to avoid the harm of false-negative labels.

\par Though advances in dealing with imperfect datasets have been made, the problem of dealing with datasets having partly missing labels that are additionally noisy in object detection tasks remains.

\par 
We propose SISSI (Seamless Iterative Semi-Supervised correction of Imperfect labels) for training object detection models with noisy and missing labels in a semi-supervised fashion, see Fig.~\ref{fig:sissi}. 
We perform several experiments with mixed-batch training, self-training with iterative label correction, synthetic-like image generation, and altering the starting point of self-training (ADELE vs. validation loss).


\begin{figure}[t]
\centering
    \begin{tabularx}{\textwidth}{   
        >{\raggedleft}p{0.24\textwidth}
        >{\raggedleft}p{0.24\textwidth}
        |
        >{\raggedleft}p{0.24\textwidth}
        >{\raggedleft}p{0.24\textwidth}   }
\multicolumn{2}{c}{\textbf{CellPose (Source)}}& \multicolumn{2}{c}{\textbf{CellLab (Target)}}
\\
\includegraphics[width=.24\textwidth]{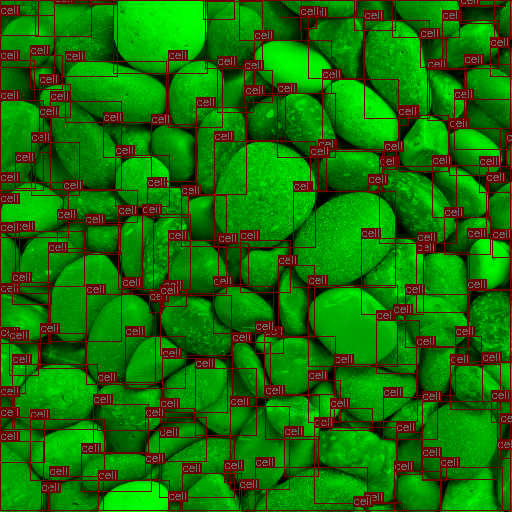}&
\includegraphics[width=.24\textwidth]{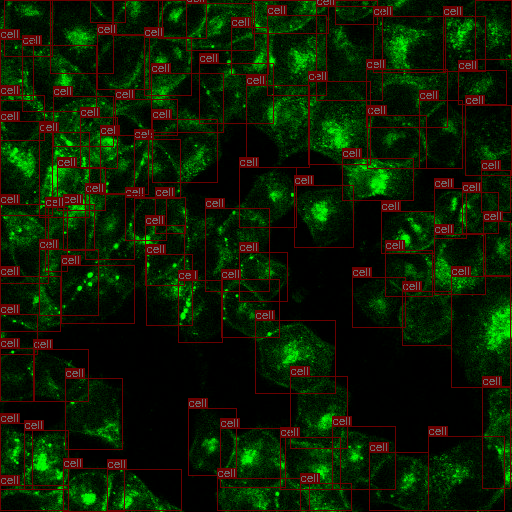}&
\includegraphics[width=.24\textwidth]{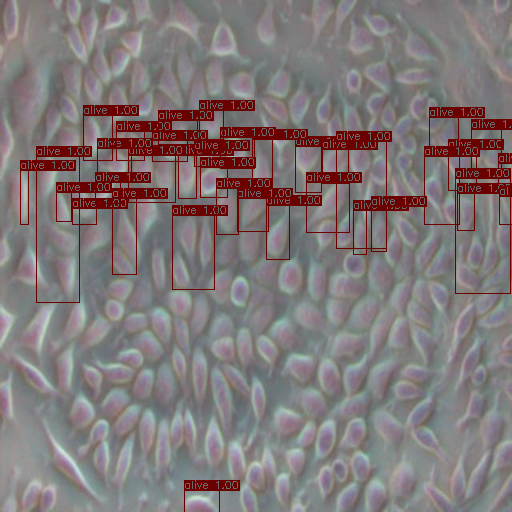}&
\includegraphics[width=.24\textwidth]{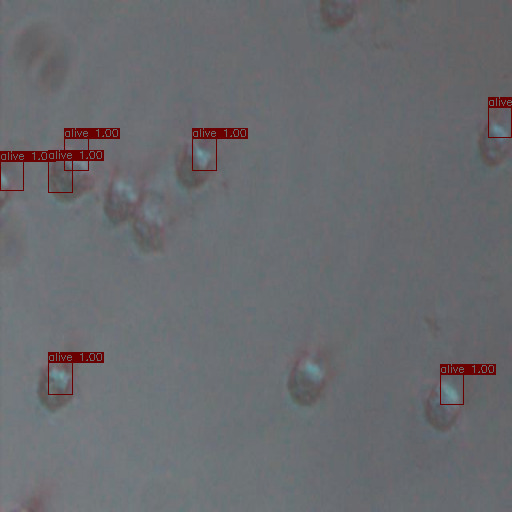}
    \end{tabularx}
\caption{Examples of the the labelled source and the noisy labelled target datasets.}
\label{fig:datasetComparison}
\end{figure}
\par 
\section{Materials and Methods}

\subsection{Datasets} 
Microscopy images of fibroblast (L929) were aquired 
using a Nikon Eclipse TS 100 microscope and the OPTOCAM-I camera.
This trainset (\textbf{CellLab} dataset) consists of 224 images, and their noisy annotations are generated with simple image processing pipelines such as Circle Hough Transform, Watershed, and Edge Detection. A detailed description of the initial weak label generation is shown in Fig.~5 in Appendix A.
The CellLab testset consists of five images $(640 \times 480)$ annotated by three cell experts. Three readers annotated five images independently, resulting in (reader 1) 552, (reader 2) 565, and (reader 3) 477 annotated cells for the five images. In order to perform domain adaptation and enhance our weak and noisy labelled CellLab dataset, we use the labelled \textbf{Cellpose}~\cite{Stringer2022} dataset.
It consists of a large variety of fluorescent markers and image modalities, as well as natural images that can be segmented into repetitive structures/blobs. The Cellpose dataset is used for training (45,215 cells on 539 images) and validation (7,195 cells on 68 images). We extract bounding boxes from the segmentation masks for our detection task. We show examples of both datasets in Fig.~\ref{fig:datasetComparison}.

\subsection{Overall Framework}
SISSI integrates a range of image processing and deep learning methods to make iterative label correction possible.

The \textbf{early learning phase} consists of a mixed-batch training combining the CellLab and Cellpose training datasets. We train the Faster R-CNN model in a supervised fashion with a Balanced Gradient Contribution \cite{Ros2016}, mixed-batch training, of target dataset with initial noisy annotations and source dataset until a memorisation phase on the noisy annotations is reached. We determine the end of early learning with a deceleration point based on the $AP_{50}$ curve between the weak ground truth of the CellLab dataset and the model output.

In the following \textbf{semi-supervised phase} for each cycle, first, we apply label correction, followed by mixed-batch training with the pseudo labels and synthetic-like images (excluding undetected cells) of the CellLab dataset combined with the original Cellpose dataset. Pseudo-label generation uses test-time augmentation and weighted boxes fusion to generate confident bounding boxes. Since some cells are not detected, their appearance in the original image will confuse the network while training. Thus, we generate dynamically synthetic-like images for continual training. The overall scheme of SISSI framework is shown in Fig.~\ref{fig:sissi}.

\subsection{Determining the Start of the Semi-Supervised Phase}

%

While training with mixed-batch training, we notice a two-stage learning phenomenon previously noted in classification and semantic segmentation: in an early learning phase, the network fits the clean annotations; then, the network start memorising the initial noisy annotations \cite{liu2020early,Liu2022adaptive}. To find the optimal point that represents when the memorisation phase starts, we adopt a method, ADELE ~\cite{Liu2022adaptive}, that has been used in previous works in the context of semantic segmentation. In our work, we rely on the deceleration of the \textit{$AP_{50}$} training curve of the model output and the initial noisy annotated dataset, CellLab, to decide when to stop trusting the initial noisy annotations and generate pseudo labels. See Fig.~7 in Appendix B for the \textit{$AP_{50}$} training curve with the point representing when the memorisation phase starts.

\subsection{Pseudo Label Generation}
Pseudo label generation is a technique where a pre-trained neural network generates labels for unlabelled data or updates labels for noisy labeled data \cite{Triguero2015self}. We generate pseudo labels to update the noisy annotations of the CellLab dataset during the semi-supervised phase.
Self-training networks have the disadvantage of being unable to correct their own mistakes. Therefore biased and wrong labels can be amplified. To filter potential bounding boxes, we integrate two techniques, test-time augmentation~(TTA)~\cite{Wang2019tta} and weighted boxes fusion~(WBF)~\cite{Solovyev2021weighted}. We average predictions generated with TTA while considering the confidence score of each bounding box in a WBF manner:
\begin{equation}
X_{1,2}=\frac{\sum ^{T} _{i=1} C_{i} \cdot X_{1,2_{i}} }{\sum ^{T} _{i=1} C_{i}},
\end{equation}
where $T$ is the number for bounding boxes assigned to a single object in a cluster, $X_{1,2}$ (or $Y_{1,2}$) is the average start and end point on the x (or y) axis. This yields the average of the bounding box coordinates $X_{1,2_{i}}$ (or $Y_{1,2_{i}}$), weighted with the confidence score $C_{i}$ for each bounding box. 

\begin{figure*}[!t]
\centering
    \begin{tabular}{c c c c c c}
         Original &\emph{t} & \emph{t+3} & \emph{t+6} & \emph{t+9}\\
         \includegraphics[width=.18\textwidth]{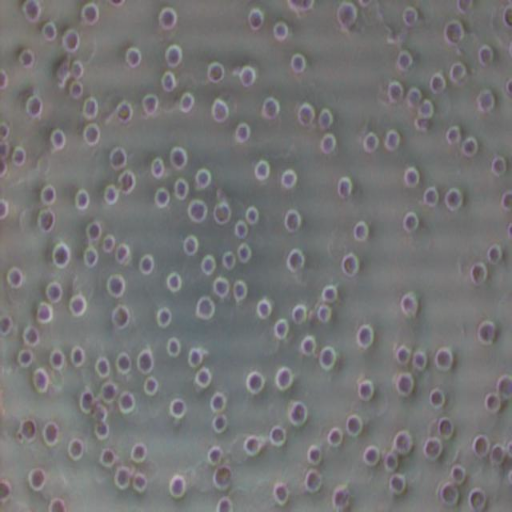}&
        \includegraphics[width=.18\textwidth]{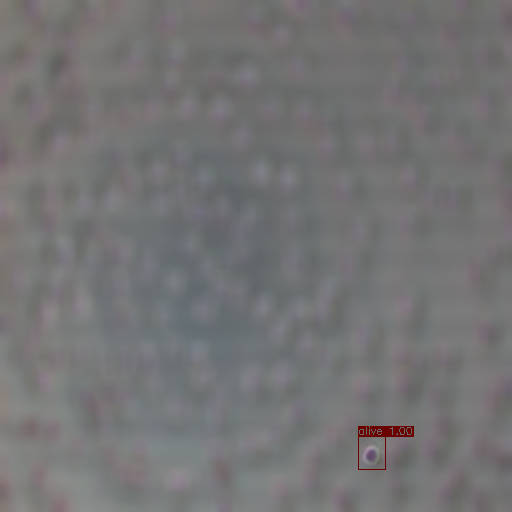} &
        \includegraphics[width=.18\textwidth]{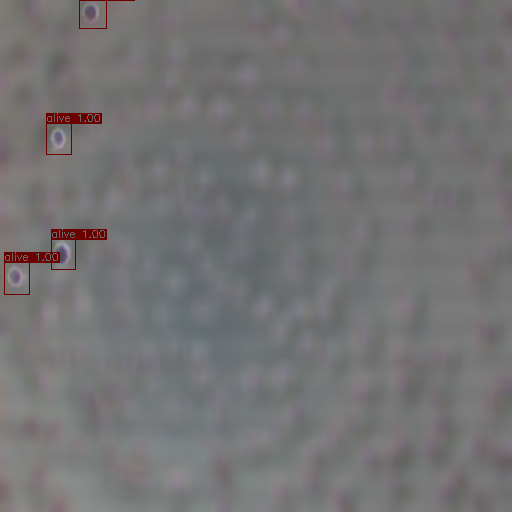} &
        \includegraphics[width=.18\textwidth]{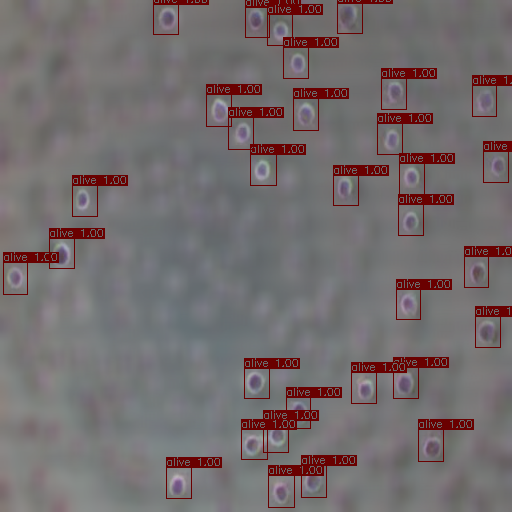} &
        \includegraphics[width=.18\textwidth]{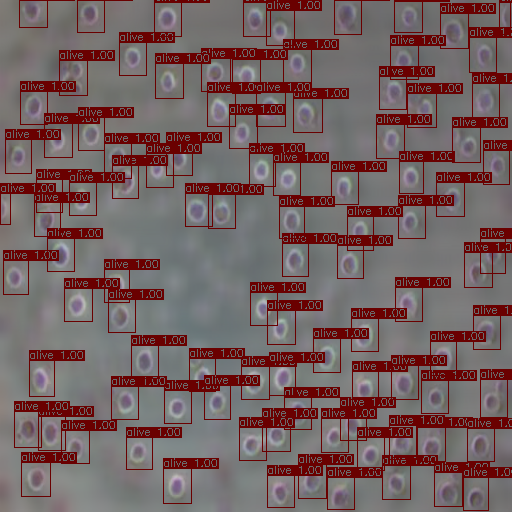}
    \end{tabular}
\caption{Example of a synthetic-like image with weak blurring in training epochs (t).}
\label{big_image_plot}

\end{figure*} 
\subsection{Synthetic-like image adaptation according to pseudo labels}
Undetected cells in the pseudo labels would affect the further training negatively. When the network localises true objects that are not present in the pseudo-labels, the network is penalised for those objects that are true. To solve this problem, we propose to generate synthetic-like images dynamically according to the pseudo labels generated for the CellLab dataset, see Fig.~\ref{big_image_plot}. To remove unlabeled cells in the training image in order not to confuse the network, we clone all the detected cells of the pseudo label (source) onto a strongly/weakly Gaussian blurred image (target). To avoid discontinuities between the target and the source, we mix edge textures with the seamless cloning algorithm (mixing gradient) \cite{Perez2003poisson}. 

\section{Experiments}

\subsection{Implementation Details} 
The backbone of our Faster R-CNN is a ResNet-50, pre-trained on the MS COCO dataset~\cite{Lin2014coco}. We set hyperparameters according to existing Fast/Faster R-CNN work \cite{Ren2015faster}. We do not freeze any layer to allow the gradient to propagate through the early layers. 

We train the models using the Stochastic Gradient Descent (SGD) optimiser with a momentum of 0.9, weight decay of 0.0002, and learning rate of 0.001. We use a batch size of 8, with an equal number of images randomly chosen from the CellLab and Cellpose datasets, and resize the images to $512\times512$. We perform simple augmentations: channel shuffle, Gaussian blurring, horizontal flip, vertical flip, and shift-scale-rotate. 
For test-time augmentation used for label correction, we use a combination of scaling ([0.8, 0.9, 1, 1.1, 1.2]) and augmentations, vertical flipping, horizontal flipping, horizontal+vertical flipping, or no flipping. We end up with 20 versions of the same image. For background blurring in the synthetic-like image generation, we use Gaussian blurring with kernels of (21, 21) and kernels (12, 32), referred to as weak(W) and strong(S) background blurring respectively.

The datasets are used as follows.
Mixed-batch training is applied in both the early supervised and semi-supervised learning phases, combining the CellLab and Cellpose training sets.
With the start of self-training, labels and synthetic-like images for the CellLab dataset are updated in each following epoch.
To perform validation for hyperparameter tuning, we use the Cellpose dataset since only five manually annotated images are available in the CellLab dataset, which all are used as a testset.
For estimating the end of early learning, the weak training labels of CellLab are compared to the model output as proposed in ADELE. We calculate deceleration by the relative change in the derivative of the \textit{$AP_{50}$} curve, and if it is above a certain threshold, 0.9, then label correction starts.
\begin{algorithm}[!t]
\caption{Pseudocode for iterative self-training with SISSI, prediction (p), target (t), bounding boxes (bbs).}\label{alg:sissi}
\begin{algorithmic}
\Require $CLimg, CLbbs_t, CPimg, CPbbs_t$ {\Comment{CellLab and Cellpose datasets}}
\Require ${NN(img)}$ \Comment{Faster R-CNN}
\Require $E \gets this.self\_training\_epoch$
\For{each pseudo\_batch $B$ in E} {\Comment{Pseudo label generation}}
\State $CLbbs_p[n], scores[n] \gets NN(TTA(CLimg\in{B}))$ \Comment{Generate boxes with TTA}
\State $CLbbs_t \gets (\sum^N_{n=1} scores[n] \cdot CLbbs_p[n])/ \sum^N_{n=1} scores[n]$ {\Comment{Filter bbs with WBF}}
\State $update\_dataset(CLbbs_t)$ {\Comment{Update final pseudo label}}
\EndFor
\For{each mixed\_batch $B$ in E} \Comment{Training}
\State $CLimg\_crops[n] \gets crop(CLimg, CLbbs)$ \Comment{Synthetic image generation}
\State $CLimg\_blur \gets blur(CLimg)$ 
\State $CLimg\_synth \gets seamless\_clone(CLimg\_blur, CLimg\_crops[n])$
\State $CLbbs_p, CPbbs_p \gets NN(CLimg\_synth, CPimg \in B)$ \Comment{Prediction}
\State $CLCP\_loss \gets loss([CLbbs_p, CPbbs_p], [CLbbs_t, CPbbs_t]) $ \Comment{Loss calculation}
\State $CLCP\_loss.backprop()$
\EndFor
\State $E.next()$
\end{algorithmic}
\end{algorithm}
\subsection{Evaluation Metrics and Results}

In Table~\ref{tab:metrics_cytonet}, we report three versions of AP, and AR over the CellLab testset. The metrics include the Pascal VOC metric ($AP_{50}$), as well as COCO evaluation metrics \cite{Lin2014coco} ($AP_{75}$, and $AP$ and $AR$ averaged over different IoU thresholds). Bold numbers denote the best performance for each of the three cell experts' annotations.

We present the detection performance of different experiments on the CellLab testset. The Baseline model is first trained in a supervised fashion with mixed-batch training of CellLab and Cellpose datasets. Aiming to correct and complete the labels, we perform early-stopping based on the validation loss of the source dataset, Cellpose, and apply self-training with test-time augmentation (TTA) and weighted boxes fusion (WBF) to iteratively update the pseudo labels. 
The following two experiments SISSI(W) and SISSI(S) additionally use synthetic-like image generation for the CellLab images, based on the pseudo labels generated with TTA and WBF. The weak (W) background blurring achieved better results than the baseline, while strong (S) background blurring has worse performance.
Plain A, is an additional experiment similar to baseline setting but with an additional algorithm (ADELE) to find an optimal starting point for label correction. This shows an increase of about 10\% across all readers in the AR metrics, compared to the baseline model, while the AP is lower.

Two versions of the final pipeline show the best results. It combines all steps (1) supervised learning with mixed-batch training of CellLab and Cellpose till a memorisation point is reached, (2) iteratively applying pseudo-label generation for the CellLab dataset with test-time augmentation and weighted boxes fusion, (3) generation of synthetic-like images according to the pseudo labels, and (4) training the network for an Epoch with mixed-batch training of the Cellpose dataset and the pseudo labels and synthetic-like images of the CellLab dataset. Pseudo code for the loop of 2-4 can be seen in Algorithm~\ref{alg:sissi}.
\begin{figure}[!t]
\centering
    \begin{tabular}{c c c c c}
        \textbf{Input} 
        &\textbf{Ground Truth}  
        &\textbf{Weak Labels}
        &\textbf{ADELE}
        &\textbf{ADELE+SISSI} \\
        \includegraphics[width=.18\textwidth]{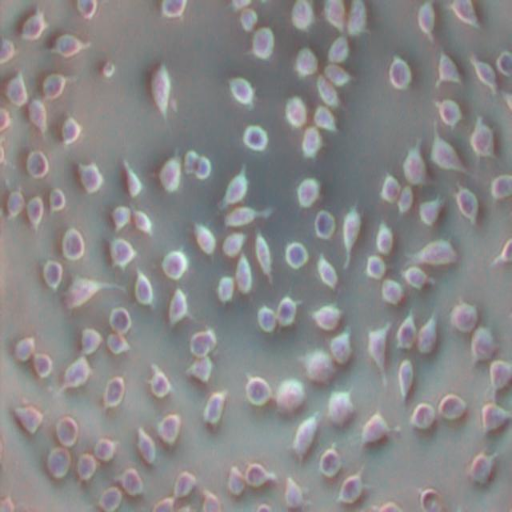}&
        \includegraphics[width=.18\textwidth]{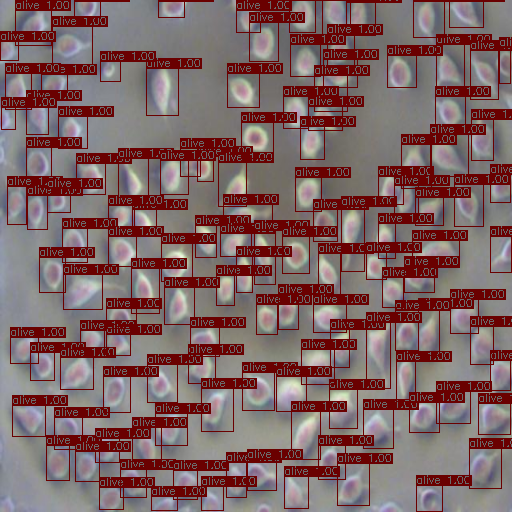}&
        \includegraphics[width=.18\textwidth]{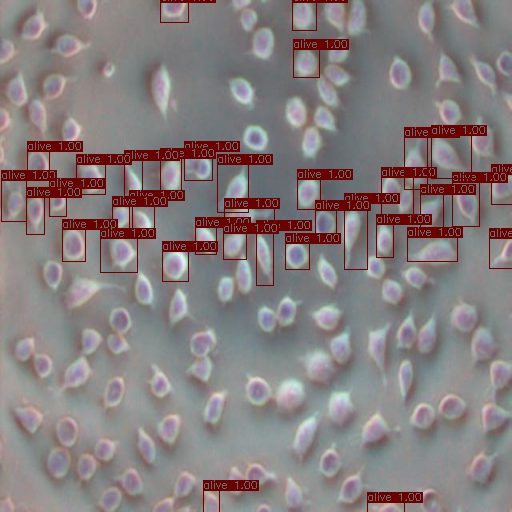} &
        \includegraphics[width=.18\textwidth]{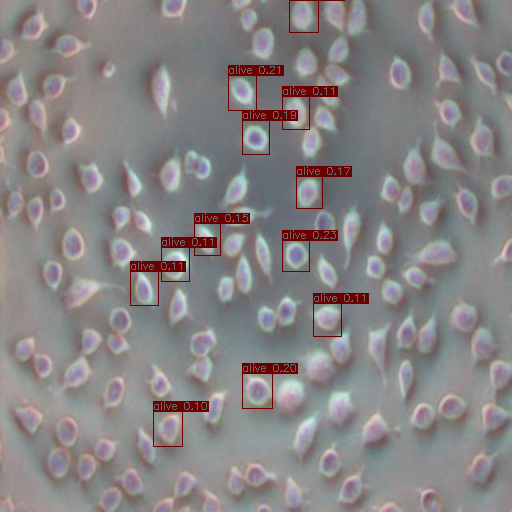} &
        \includegraphics[width=.18\textwidth]{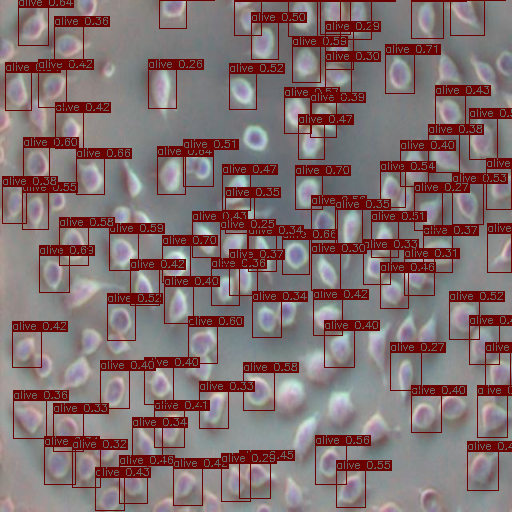}
    \end{tabular}
\caption{Demonstration of improvement of results with our proposed method.}
\label{fig:visual_perforamnce}
\end{figure} 
Incorporating ADELE with our label correction and synthetic-like image generation method with strong blurring increases the AP by at least 15\% and AR by at least 20\% compared to the baseline across all cell experts. On~Fig.~\ref{fig:visual_perforamnce}, we show an example where SISSI successfully
improves the detection results of Plain A (ADELE) experiment. Examples of a training image with its pseudo labels for different epochs (t) and
experiments can be seen in Fig.~8 in Appendix B.

\begin{table*}[!h]
    \centering
    \caption{Results of Cell Detection on the CellLab testset.}
    \label{tab:metrics_cytonet}
    \begin{tabularx}{\textwidth}
    {   >{\raggedright}p{0.16\textwidth}|
        >{\raggedleft}p{0.06\textwidth}
        >{\raggedleft}p{0.06\textwidth}
        >{\raggedleft}p{0.06\textwidth}
        >{\raggedleft}p{0.06\textwidth}
        |
        >{\raggedleft}p{0.06\textwidth}
        >{\raggedleft}p{0.06\textwidth}
        >{\raggedleft}p{0.06\textwidth}
        >{\raggedleft}p{0.06\textwidth}
        |
        >{\raggedleft}p{0.06\textwidth}
        >{\raggedleft}p{0.06\textwidth}
        >{\raggedleft}p{0.06\textwidth}
        >{\raggedleft\arraybackslash}p{0.06\textwidth}   }
& \multicolumn{4}{c|}{Annotator 1} &\multicolumn{4}{c|}{Annotator 2}&\multicolumn{4}{c}{Annotator 3}
\\\hline
Pipeline    
 & $AP_{50}$ & $AP_{75}$&  $AP$& $AR$
 & $AP_{50}$ & $AP_{75}$&  $AP$& $AR$
 & $AP_{50}$ & $AP_{75}$& $AP$& $AR$
\\\hline\hline
Baseline
&   45.6	&15.7   &21.3	&32.7
&   44.3	&16.1   &20.2	&31.0
&   \textbf{58.6}	&28.1   &29.7	&41.8
\\\hline 
SISSI(W)       
&	52.8&	23.2&   25.9&	37.7
&	49.5&	21.3&   24.4&	34.6
&	58.5&	29.0&   29.7&	42.0
\\\hline
SISSI(S)      
&	40.3&	8.3&    16.2&	32.8
&	38.4&	7.2&    15.2&	31.8
&	46.6&	9.7&    19.2&	37.8
\\\hline
Plain A                       
	&38.7	&18.9&19.2	&43.9
	&35.8	&18.6&18.7	&43.1
	&41.2	&23.8&22.9	&50.6
\\\hline
A+SISSI(W)
&	43.1&	37.1&   36.0&	\textbf{60.3}
&	45.1&	38.5&   37.4&	\textbf{58.8}
&	47.6&	42.8&   41.4&	\textbf{66.9}
\\\hline
A+SISSI(S)
&	\textbf{54.9}
&	\textbf{49.0}
&   \textbf{43.2}
&	57.6
&	\textbf{51.2}
&	\textbf{45.1}
&   \textbf{39.7}
&	54.3
&	58.5
&	\textbf{55.5}
&   \textbf{47.9}
&   64.9
\\\hline
\end{tabularx}
\end{table*}

\section{Discussion}

\subsection{Findings}
The experiments made clear that both the start of label correction and the amount of background information appearing in images during training impact the results.
When starting label correction too early, during early learning, the network is not confident enough to detect all objects in the image; thus, correcting initially noisy annotations at this stage results in a high rate of missing targets. Training a network on images with high missing targets without SISSI (Plain A) increases the uncertainty of the network compared to label correction in a later memorization phase with fewer missing targets, the baseline.
\par
In the basic SISSI approach, where label correction is started on a model chosen based on the validation loss of the external Cellpose dataset, weak background blurring worked better than a strongly blurred background. We believe this phenomenon appears because the neural network has learned more contextual information in the memorisation stage and requires the background information.

On the other hand, starting label correction when the early learning phase ends, according to ADELE, strong blurring shows better results than weak blurring. The information about the background is less important. This can be an advantage in synthetic-like image generation because figuring out how to preserve contextual information seems less critical.

\subsection{Limitations}
The success of SISSI may be dependent on the stopping criteria and the training phase, early learning/memorisation phase. When the annotations in the image are too noisy, the network may not encompass the early learning phase as in previous works, ADELE. It may be unable to learn the task to produce new pseudo labels for further training. The effect of blurring during different training phases needs more empirical research for verification. SISSI is a simple approach that works with only one class of interest to detect. Blurring with multi-object needs further modification in future works. We use SISSI in these experiments with Faster R-CNN, which is more robust and friendly for the missing label scenario than other detection networks.

\subsection{Conclusion}
This paper presents a method to train object detection models with noisy and missing annotations with semi-supervised learning by proposing a novel technique. We use dynamically generated synthetic-like images using seamless cloning for further training the network after pseudo-label generation. We utilize a domain adaptation technique, Balanced Gradient Contribution, to generate stable gradient directions and mitigate the so noisy annotation problem for our semi-supervised training. Finally, we evaluate our method for the cell detection task with various training procedures and show its improvement over the usual semi-supervised approach. Our method, SISSI, can be added on top of any detection network, and it also helps other methods like ADELE to be leveraged for object detection.
In the future, we will adapt our method to work with multi-object detection and explore SISSI with different detection networks. Moreover, we will explore our method for different medical detection tasks and integrate our network to help cell experts with the grading task.

\subsubsection{Acknowledgements}
We would like to acknowledge Oesterreichisches Forschungsinstitut für Chemie und Technik~(OFI) for CellLab images and test set annotation. We would also like to thank Dr. Thomas Mohr for revising the manuscript.


%
%
%
\bibliographystyle{splncs04}
\bibliography{SISSI}

\begin{thebibliography}{10}
\providecommand{\url}[1]{\texttt{#1}}
\providecommand{\urlprefix}{URL }
\providecommand{\doi}[1]{https://doi.org/#1}

\bibitem{Anderson2016iso}
Anderson, J.M.: { Future challenges in the in vitro and in vivo evaluation of
  biomaterial biocompatibility }. Regenerative Biomaterials  \textbf{3}(2),
  73--77 (03 2016). \doi{10.1093/rb/rbw001},
  \url{https://doi.org/10.1093/rb/rbw001}

\bibitem{Gao_2019_ICCV}
Gao, J., Wang, J., Dai, S., Li, L.J., Nevatia, R.: Note-rcnn: Noise tolerant
  ensemble rcnn for semi-supervised object detection. In: Proceedings of the
  IEEE/CVF International Conference on Computer Vision (ICCV) (October 2019)

\bibitem{International2009iso}
ISO: Iso 10993-5: 2009-biological evaluation of medical devices-part 5: Tests
  for in vitro cytotoxicity (2009)

\bibitem{Li2022}
Li, H., Pan, X., Yan, K., Tang, F., Zheng, W.S.: Siod: Single instance
  annotated per category per image for object detection. In: Proceedings of the
  IEEE/CVF Conference on Computer Vision and Pattern Recognition (CVPR). pp.
  14197--14206 (2022)

\bibitem{Lin2014coco}
Lin, T.Y., Maire, M., Belongie, S., Bourdev, L., Girshick, R., Hays, J.,
  Perona, P., Ramanan, D., Zitnick, C.L., Dollár, P.: Microsoft coco: Common
  objects in context (2014), \url{http://arxiv.org/abs/1405.0312}

\bibitem{Liu2022adaptive}
Liu, S., Liu, K., Zhu, W., Shen, Y., Fernandez-Granda, C.: Adaptive
  early-learning correction for segmentation from noisy annotations. In:
  Proceedings of the IEEE/CVF Conference on Computer Vision and Pattern
  Recognition. pp. 2606--2616 (2022)

\bibitem{liu2020early}
Liu, S., Niles-Weed, J., Razavian, N., Fernandez-Granda, C.: Early-learning
  regularization prevents memorization of noisy labels. Advances in Neural
  Information Processing Systems  \textbf{33} (2020)

\bibitem{Lyu2021}
Lyu, F., Yang, B., Ma, A.J., Yuen, P.C.: A segmentation-assisted model for
  universal lesion detection with partial labels. In: de~Bruijne, M., Cattin,
  P.C., Cotin, S., Padoy, N., Speidel, S., Zheng, Y., Essert, C. (eds.) Medical
  Image Computing and Computer Assisted Intervention -- MICCAI 2021. pp.
  117--127. Springer International Publishing, Cham (2021)

\bibitem{Perez2003poisson}
P\'{e}rez, P., Gangnet, M., Blake, A.: Poisson image editing. ACM Trans. Graph.
   \textbf{22}(3),  313–318 (jul 2003). \doi{10.1145/882262.882269},
  \url{https://doi.org/10.1145/882262.882269}

\bibitem{Ren2015faster}
Ren, S., He, K., Girshick, R., Sun, J.: Faster r-cnn: Towards real-time object
  detection with region proposal networks. Advances in neural information
  processing systems  \textbf{28} (2015)

\bibitem{Ros2016}
Ros, G., Stent, S., Fernández~Alcantarilla, P., Watanabe, T.: Training
  constrained deconvolutional networks for road scene semantic segmentation.
  CoRR  (04 2016)

\bibitem{Siegismund2018developing}
Siegismund, D., Tolkachev, V., Heyse, S., Sick, B., Duerr, O., Steigele, S.:
  Developing deep learning applications for life science and pharma industry.
  Drug research  \textbf{68}(06),  305--310 (2018)

\bibitem{Solovyev2021weighted}
Solovyev, R., Wang, W., Gabruseva, T.: Weighted boxes fusion: Ensembling boxes
  from different object detection models. Image and Vision Computing
  \textbf{107},  104117 (2021)

\bibitem{Stringer2022}
Stringer, C., Pachitariu, M.: Cellpose 2.0: how to train your own model.
  bioRxiv  (2022). \doi{10.1101/2022.04.01.486764},
  \url{https://www.biorxiv.org/content/early/2022/04/05/2022.04.01.486764}

\bibitem{Tajbakhsh2020}
Tajbakhsh, N., Jeyaseelan, L., Li, Q., Chiang, J.N., Wu, Z., Ding, X.:
  Embracing imperfect datasets: A review of deep learning solutions for medical
  image segmentation. Medical Image Analysis  \textbf{63},  101693 (2020).
  \doi{https://doi.org/10.1016/j.media.2020.101693},
  \url{https://www.sciencedirect.com/science/article/pii/S136184152030058X}

\bibitem{Triguero2015self}
Triguero, I., Garc{\'\i}a, S., Herrera, F.: Self-labeled techniques for
  semi-supervised learning: taxonomy, software and empirical study. Knowledge
  and Information systems  \textbf{42}(2),  245--284 (2015)

\bibitem{Wang2019tta}
Wang, G., Li, W., Aertsen, M., Deprest, J., Ourselin, S., Vercauteren, T.:
  Aleatoric uncertainty estimation with test-time augmentation for medical
  image segmentation with convolutional neural networks. Neurocomputing
  \textbf{338},  34--45 (2019)

\bibitem{Wang2022unreliable}
Wang, Y., Wang, H., Shen, Y., Fei, J., Li, W., Jin, G., Wu, L., Zhao, R., Le,
  X.: Semi-supervised semantic segmentation using unreliable pseudo-labels. In:
  Proceedings of the IEEE/CVF Conference on Computer Vision and Pattern
  Recognition (CVPR). pp. 4248--4257 (June 2022)

\bibitem{Xiong2022UnetCHT}
Xiong, H., Liu, S., Sharan, R.V., Coiera, E., Berkovsky, S.: Weak label based
  bayesian u-net for optic disc segmentation in fundus images. Artificial
  Intelligence in Medicine  \textbf{126},  102261 (2022)

\bibitem{Zhang2020recalibrationloss}
Zhang, H., Chen, F., Shen, Z., Hao, Q., Zhu, C., Savvides, M.: Solving
  missing-annotation object detection with background recalibration loss.
  ICASSP 2020 - 2020 IEEE International Conference on Acoustics, Speech and
  Signal Processing (ICASSP) pp. 1888--1892 (2020)

\end{thebibliography}

\end{document}


\maketitle

\appendix
\setcounter{figure}{4}    
\section{Initial Weak Labels Generation With Image Processing.}
\begin{figure}[!htb]
\centering
    \begin{tabular}{c c}
        \includegraphics[width=.34\textwidth]{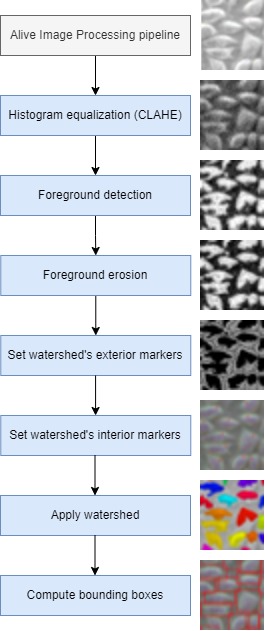}&
        \includegraphics[width=.34\textwidth]{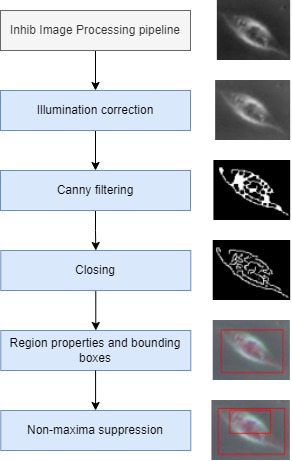}\\
    \end{tabular}
\label{ip_pipelines}
\caption{Cell Images are divided visually and manually into 3 categories (alive/inhibited/dead) to perform suitable image processing algorithms to generate initial weak labels. Alive and inhibited pipelines are shown in this Figure.}
\end{figure}
\clearpage

\begin{figure}[!htb]
    \centering
    \includegraphics[width=.5\textwidth]{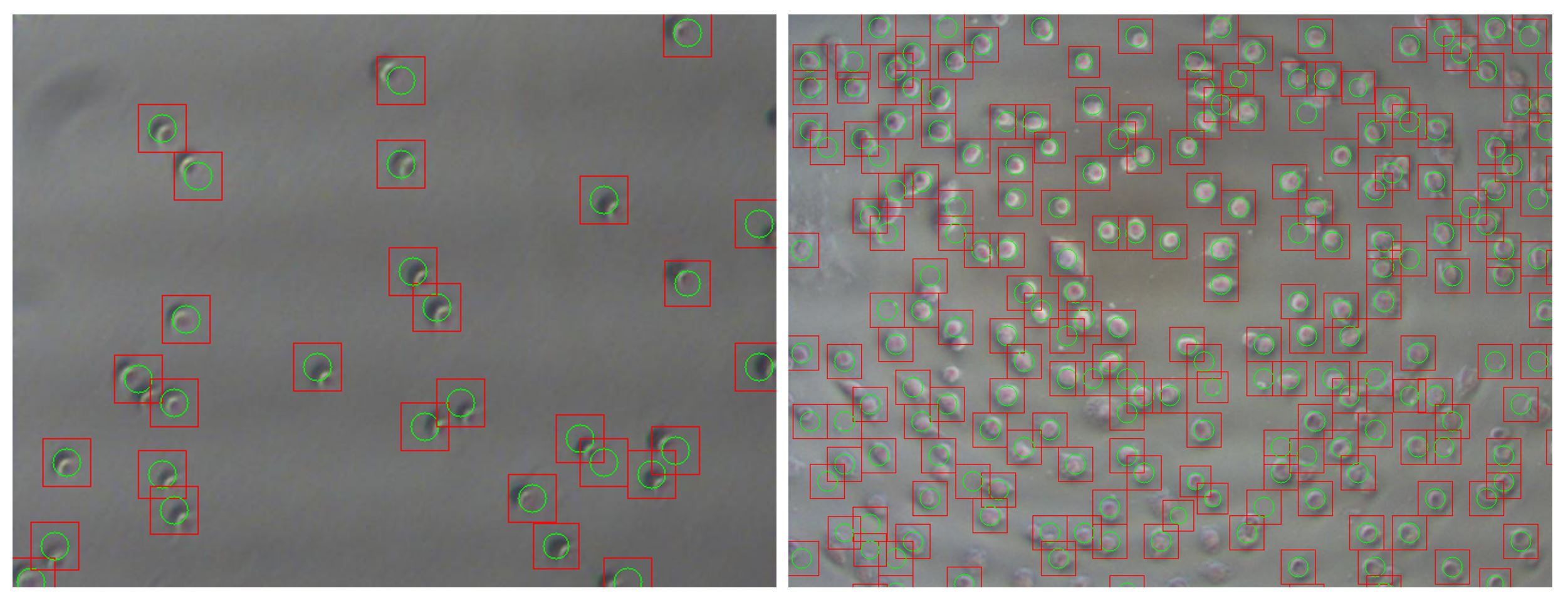}
    \caption{We perform Circle Hough Transform directly on the dead cells image.}
    \label{fig:dead_cell_bb}
\end{figure}

\section{SISSI Training Dynamics.}
\begin{figure}[!htb]
    \centering
    \includegraphics[width=.5\textwidth]{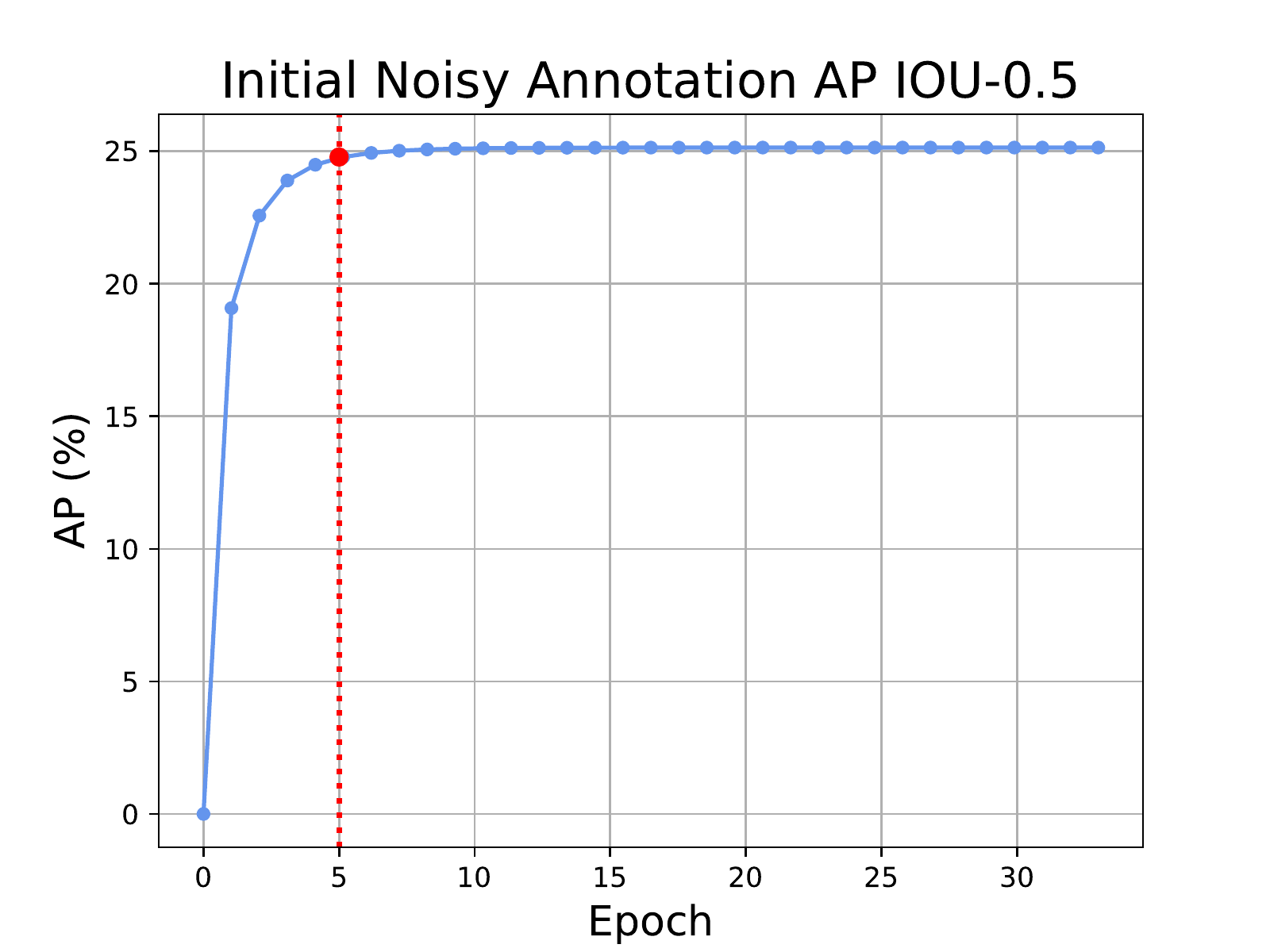}
    \caption{Fitted IoU curve between the model output and the initial noisy annotations with the point where we start label correction based on the $AP_50$ curve deceleration.}
    \label{fig:adele_curve}
\end{figure}

\begin{figure*}[!htb]
\centering
    \begin{tabular}{M{2.5cm}M{2.5cm}M{2.5cm}M{2.5cm}M{2.5cm}}
         & \emph{t} & \emph{t+3} & \emph{t+6} & \emph{t+9}\\
        \textbf{Plain A} &
        \includegraphics[width=.15\textwidth]{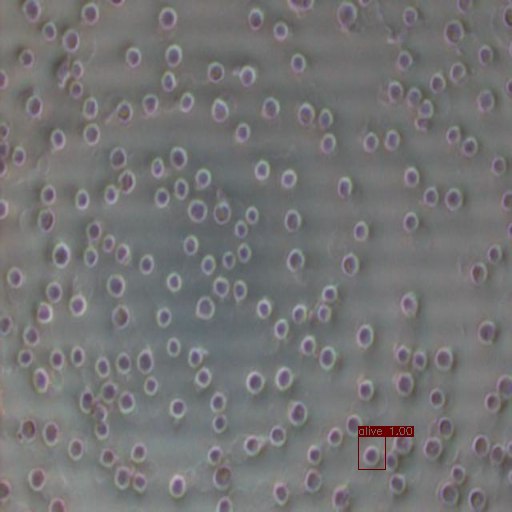} &
        \includegraphics[width=.15\textwidth]{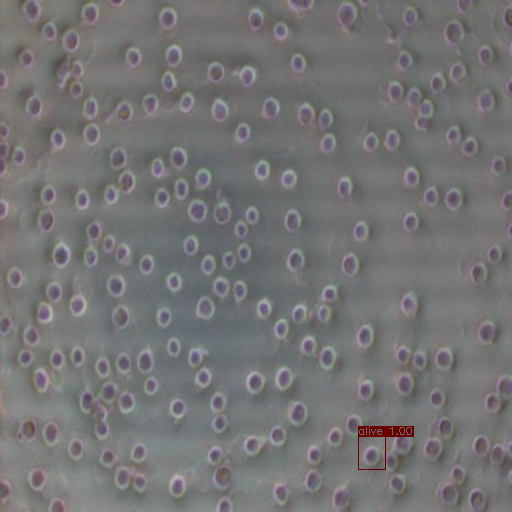} &
        \includegraphics[width=.15\textwidth]{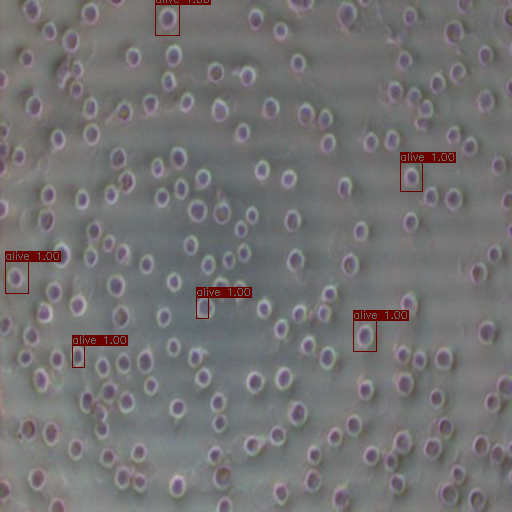} &
        \includegraphics[width=.15\textwidth]{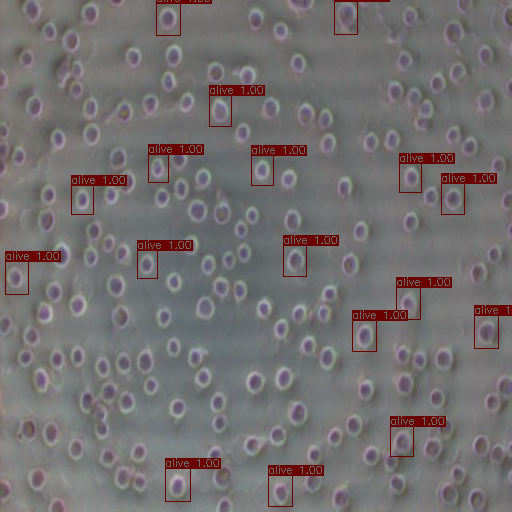}\\
        
        \textbf{A+SISSI (W)} &
        \includegraphics[width=.15\textwidth]{graphics/training_over_time/A+SISSI/epoch5.png} &
        \includegraphics[width=.15\textwidth]{graphics/training_over_time/A+SISSI//epoch7.png} &
        \includegraphics[width=.15\textwidth]{graphics/training_over_time/A+SISSI//epoch9.png} &
        \includegraphics[width=.15\textwidth]{graphics/training_over_time/A+SISSI//epoch12.png}\\

    \end{tabular}

\caption{Examples of a training image with its pseudo labels for different epochs (t) and experiments.}
\label{big_image_plot}

\end{figure*} 